\documentstyle[12pt,psfig]{article}
\begin{document}
\date{}
%\draft
\bibliographystyle{unsrt}
%\hfil{\today}
%\begin{title}
\title{A New Algorithm for Protein Design}
%\end{title}
\author{J.M.Deutsch and Tanya Kurosky\\
University of California, Santa Cruz, U.S.A.}
\maketitle
%\noreportnumber
\abstract{We apply a new approach to the reverse protein folding problem.
Our method uses a minimization function in the design process which is
different from the energy function used for folding.  
For a lattice model, we show that this new approach produces
sequences that are likely to fold into desired structures.  Our
method is a significant improvement over previous attempts which
used the energy function for designing sequences.}

\newpage

\section{Introduction}

A problem of current interest is the ``reverse problem''
for protein folding: given a desired structure with
coordinates $\Gamma \equiv \{  r_1,\dots,r_N \}$, what is the
best choice of sequence $S$ to obtain that structure? 
This problem is of great importance to
biotechnology as it facilitates the design of new drugs.
Progress towards a solution of this problem has been made\cite{yue&dill}, but
there is still no trustworthy algorithm to accomplish
this task. 

To find the
optimum sequence it would seem necessary to apply
some minimization procedure over the space of all
possible sequences $S$. This had previously been
attempted\cite{SG,shakhnovich} by taking the Hamiltonian, which depends
both on $\Gamma$ and $S$ and minimizing it over $S$,
at fixed $\Gamma$. We will discuss this in more detail
below. As was recently argued\cite{us}, this is not guaranteed
to work, and in general is not an optimal design strategy.
Instead a new method was introduced that provides a
systematic approach to this problem. It was shown
that a minimization function exists though it
is more numerically intensive to evaluate than the
Hamiltonian. Applied to a system of block copolymers,
the method proved to be highly successful\cite{us}.

The purpose of this paper is to examine this method
for some simple models of much recent interest, the 27-mer
cube with ising interactions\cite{SG}, 
and the two dimensional HP model of Lau and Dill\cite{HP}. 
It will be shown below that this new
minimization function works very well and gives results 
far better than the minimization of the Hamiltonian attempted
previously\cite{SG}.

In designing proteins, three criteria are generally
considered to mark success.  
First, the engineered molecule should have
the desired conformation when it is in its ground state.
Second, there should be no ground state degeneracy, so that the
molecule will always fold into the same conformation.  Third,
there should be a large gap in the energy of the ground state
and the energies of the higher energy states so that the 
molecule will stably sit in the ground state.   
For some systems the method of minimizing the Hamiltonian can be
shown to always be successful in finding a sequence which has
a desired conformation as its ground state, as we will
examine below. But
sadly, we will see that the 
sequences designed this way will usually fold up into thousands of other
structures, all with the identical energy. 
Our new method gives sequences with a much lower
degeneracy, often producing unique ground states in the 
desired conformation, and failing quite rarely.  
In addition, we predict that it will
also result in sequences with a large gap in the energy 
spectrum above the ground state.  

This work is significant in that 
it describes a general method to apply to any 
model of protein folding and is
based on a systematic approach to the problem. Previous work\cite{SG}
does not appear to have recognized the need for a less
{\em ad hoc} approach, and instead blamed the nature
of the problem, rather than the faults in its solution.
For example, it has been pointed out\cite{whoops} 
that within a given model there are ``good" desired
structures, structures that have  sequences that will fold uniquely to
them. Structures that do not have this property are deemed ``bad".
When the method of minimizing the Hamiltonian failed to find unique
ground states, it was argued that it was because the 
initial structures chosen were ``bad"
structures.  As a result, it has been thought that more complicated
models were needed so that ground state degeneracies would be
reduced\cite{whoops},\cite{80-mer}, and that some means of
distinguishing between ``good" initial structures and ``bad" ones might be
necessary\cite{whoops}.  While these considerations are valid, 
our results show that the basic
problem of how to find the sequence with the lowest degeneracy that
folds into a given random conformation has not been addressed properly 
in previous work. For a model where we can quantify
the success of this previous method, we find that most 
``good" structures are misdesigned.
It is therefore important to fully understand the basic problem
of reverse folding before moving on to more complicated models.  In
this letter we hope to clarify the basic logic behind the reverse folding
problem.  

The outline for this letter is as follows. First the
results of this new method are summarized. Then the
approximate but numerically efficient scheme that we use is
briefly reviewed and related to recent work. In section 
\ref{sec:results} this
method is applied numerically to the case of two
different molecular types on a $3 \times 3 \times 3$ lattice. 
It is also compared with
the previous method of Shakhnovich and Gutin. 
We then do a numerical comparison of the two dimensional
HP model with 16 monomers, where an even more detailed
comparison can be made.  Finally the feasibility of
other extensions to the this work is discussed.

\section{Minimization Function}
\label{sec:minimization}

Here we summarize previous results\cite{us} in 
which we obtained the minimization function 
appropriate for use in the reverse folding problem.  
Given a system with coordinates denoted by $\Gamma$ and chemical sequence
by $S$, the probability
that a sequence $S$ gives the desired structure $struct$ is

\begin{equation}
P(S|struct) \propto\exp ({-(F_{struct}(S)-F_o(S))/T})\equiv \exp ({-\Delta F/T})
\label{eq:minimize}
\end{equation}
where
\begin{equation}
\exp ({-F_o(S)/T}) \equiv \sum_{\Gamma} \exp ({-H_S(\Gamma)/T})
\end{equation}
\begin{equation}
\exp ({-F_{struct}(S)/T}) \equiv \sum_{\Gamma} \exp ({-(H_S(\Gamma)+V_{ext}(\Gamma))/T}) ~,
\end{equation}
with the sums being done over all possible structures.
 
$F_{struct}(S)$ is the free energy {\it pushed}
into a certain structure by the ``clamping potential", $V_{ext}(\Gamma)$, which 
is chosen to force the molecule to have the
desired structure.  So the optimum sequence is the one with the smallest
difference $\Delta F$ between the unrestricted free
energy and the free energy ``clamped" in the desired
structure. 

This can be understood physically as follows. Suppose one desired
to create a protein  of $N$ amino acids that folded into a 
desired structure $struct$.  One could imagine doing this in a brute force 
way by creating all $20^N$ possible polypeptide chains all
maintained at equilibrium in a container. This is the ``sequence
space soup" of Chan and Dill\cite{soup}.  One wishes to pick
out the molecule most closely resembling the desired structure.
To do this, one creates a kind of ``fish hook", that is a scaffolding
potential for which only molecules resembling $struct$ will
fit. This is the same as $V_{ext}$.  $V_{ext}$, by design, 
will predominantly 
``catch" molecules of the
desired sort. In fact the probability of it containing a molecule
with sequence $S$ is precisely the same as eqn. (\ref{eq:minimize}).

It is also apparent that by minimizing the  $\Delta F$ of 
eqn. (\ref{eq:minimize}), one is not only finding a sequence
that will fold into the correct structure, but also the
sequence that has the highest probability of being in the
desired structure, at finite temperature. Therefore this
method satisfies the the third criterion stated above.
The above method has much in common with those used in
learning theory of ``neural networks''\cite{hertz}.

\section{Application to lattice models}
We look at two simplified lattice models with a self-avoiding chain and
apply our design algorithm to it. The first model has been studied much
recently in connection with the design problem\cite{SG}.  
We use a $3\times 3 \times 3$ lattice and a 27-monomer chain so that it is 
possible to directly enumerate the energies of all 
conformations with a given sequence. Thus we can easily evaluate
the efficacy of our method.
 
The model involves sequences
$\{ \sigma_i\}$ of
two possible monomer types that are given values $\pm 1$, for chains of length
$N$. The monomer type values, $\{ \sigma_i\}$, plus the
positions of all the monomers, $\{r_i\}$, completely describe the state
of the chain.  The energy is
\begin{equation} 
E(\{\sigma_i\}, \{r_i\}) = 
{1\over 2} \sum_{i,j}^N (B_0 + B\sigma_i\sigma_j)\Delta({\bf r}_i-{\bf r}_j)
\label{eq:model}
\end{equation}
We have taken $\Delta({\bf r}_i-{\bf r}_j) = 1$ if $\{r_i\}$ and $\{r_j\}$ 
are nearest neighbors and zero otherwise.
$B$ is negative so that the monomer types will segregate, giving 
ferromagnetic ordering of the $\sigma$'s.  $B_0$ has been taken to
have a large magnitude and is negative
so that there is a large attraction between all nearest neighbors,
causing the protein to collapse into a shape of minimal surface
area, in this case a cube. Thus the only conformations we
need consider are the internal arrangements of a chain packed
into a cube.

To find the sequence most likely to fold into some desired conformation, 
we want to minimize $\Delta F$ as defined in 
(\ref{eq:minimize}).  Specializing to the model considered
here, the clamping potential is a delta function since we would
like to pick out one specific structure.  If we call the coordinates of the
desired structure $\{r^0_i\}$, $\Delta F$ becomes
\begin{equation} 
\Delta F = E(\{\sigma_i\}, \{r^0_i\}) - F_o(\{\sigma_i\}) 
\label{eq:correct}
\end{equation}  

As described in previous work\cite{us}, we expand out $F(\{\sigma_i\})$,
keeping only the lowest order cumulant. Neglecting constant terms, this gives
\begin{equation} 
F_o(\{\sigma_i\}) \approx {B\over 2} \sum_{i,j}^N  
\sigma_i\sigma_j\langle\Delta({\bf r}_i-{\bf r}_j)\rangle ,
%TANYA: Please keep the , where it is.
%ok
\label{eq:F_o}
\end{equation}
with the average done over all compact conformations with minimal
surface area.  Note that this term gives an anti-ferromagnetic Ising
interaction.  Thus we wish to minimize
\begin{equation} 
\Delta F \approx {B\over 2} \sum_{i,j}^N  
[\Delta({\bf r}_i-{\bf r}_j) - \langle\Delta({\bf r}_i-{\bf r}_j)\rangle]
\sigma_i\sigma_j
\label{eq:approx}
\end{equation} 
as a function of the $\{ \sigma_i\}$.

The first work on this model, by Shakhnovich and Gutin, used as
stated earlier, energy minimization. To keep the chain from becoming
a homopolymer, they added the constraint of constant magnetization.
It has met with some degree of success and we have 
previously attempted\cite{us} to
analyze why that should be in the systematic framework 
that we have recently developed.
This method can be interpreted\cite{us} as
an approximation to eqn. (\ref{eq:approx}). If we approximate
$\langle\Delta({\bf r}_i-{\bf r}_j)\rangle$ to be a constant, then
$F_o$ is equivalent to an anti-ferromagnetic mean field interaction
proportional to the total magnetization squared. This has the effect
of preferring a total magnetization close to zero, and is therefore similar
to the constraint of a fixed magnetization used by Shakhnovich and Gutin. 
It has further come to our attention that Shakhnovich and Gutin ignored
the interactions along the backbone\cite{private}. 
It is clear that in designing sequences, the energy of interaction along
the backbone of the chain cannot matter, since it does not depend on the
conformation. This is therefore a sensible design procedure. 
This is equivalent
to assuming that $\langle\Delta({\bf r}_i-{\bf r}_j)\rangle$ is unity, for
adjacent $i$ and $j$. For all other $i$ and $j$, it is the same as above,
similar to taking it to be a smaller constant. This is a 
better approximation to $\langle\Delta({\bf r}_i-{\bf r}_j)\rangle$
but is still fairly crude.
Notice that the contribution along the backbone of the chain naturally 
cancels in (\ref{eq:approx}); thus with our method there is no need to 
ignore backbone interactions.

Shakhnovich and Gutin\cite{SG} have noted that not only do we want to design
sequences which have a low ground state degeneracy, but also we want
sequences which have a significant energy gap between the native state
and higher energy states\cite{SG}. This is necessary in order to get
stable ground state structures. After designing sequences by minimizing
the energy in sequence space, they noted that their designed sequences
tended to have more of an energy gap than randomly chosen sequences.
However, they did not decide to optimize this gap in their search of
sequence space. As noted above, minimizing the exact $\Delta F$ will give
maximally stable structures. Furthermore, in the above 
approximation to $\Delta F$,
that is $\Delta F = E(\{r^0_i\}) - \langle E(\{r_i\})\rangle$, it can
also be seen to be similar to the requirement of a large energy gap.
$E(\{r^0_i\})$ is the minimum energy conformation and $\langle
E(\{r_i\})\rangle$ is the average over all conformations. Maximizing the
difference between the minimum and the average is likely to result in a
large energy gap.

\section{Numerical Results}
\label{sec:results}
For a given desired configuration, we minimized (\ref{eq:approx}) in 
sequence space using simulated annealing in order to find the 
sequence most likely to have the desired configuration as its ground state.
The second term of (\ref{eq:approx}) was calculated
by directly enumerating all possible configurations.
We carried out the above procedure 
on 2066 randomly picked initial configurations out of the total
number of 103346 distinct configurations of the 27-mer cube.  We then
folded the resulting molecules by minimizing the energy in configuration
space.  It was found that each of the resulting molecules did indeed 
have the desired configuration as its ground state.
In addition, the average 
degeneracy of these ground states is only 3.37.  Thus
the molecules we designed are quite likely to fold into the 
desired conformations.

We now compare these results with results we have obtained
by a constrained minimization of the energy 
instead of minimizing our new function $\Delta F$. 
We performed the same simulated annealing procedure for the same 
2066 given conformations, this time minimizing the energy (in sequence
space), keeping the total magnetization equal to one. This
last constraint is necessary to keep the chain from becoming a
homopolymer, and is in keeping with work done by Shakhnovich and Gutin\cite{SG}.
Although the sequences 
found to minimize the energy can be proved to always 
have the desired conformations 
as ground states, the average degeneracy of these ``best" sequences is 1155, 
averaged over 2066 random conformations!  Clearly these
sequences are not likely to be found in the desired conformations.

We also wanted to compare our results with the method of minimizing
the energy in sequence space while ignoring interactions along the
backbone of the chain.  As discussed above, this is quite similar 
to minimizing $\Delta F$, though the results are not as good.  
The sequences produced this
way were usually found to be ground states of the desired conformation, but
they have an average ground state degeneracy of 4.11. We also find that
16 of the conformations were misdesigned as they do not fold
up into the desired structure.

In addition, we examined the  2D HP model of Lau and Dill\cite{HP}.  We
looked at 16-mers on an open 2D lattice with two types of monomers,
labeled hydrophobic (H) and polar (P), with an energy of -1 assigned to 
pairs of hydrophobes that are nearest neighbors. HP and PP interactions
are both zero.  In this model
the contribution to the energy from pairs along the backbone of the
chain is generally ignored.  
We first folded up all possible sequences and from that found that
584 conformations were ``good" structures.
In other words, for each of these structures it was possible to
find at least one sequence which would fold into it 
as a unique ground state.
We then went through all the ``good" structures
and tried to design each one, using the method of Shakhnovich
and Gutin, and by using our new method.

For the method of Shakhnovich and  Gutin, 111 structures were
missed. That is, their method failed to design these sequences 
to fold into the target structure. Of the remaining ones, only 123
folded correctly to a unique ground state. 
This represents an overall success rate of 21\%.  
The new method fared much better. 
We employed an approximation analogous to eqn. (\ref{eq:F_o}) in
estimating $F_o$. For this we had to estimate the
monomer-monomer correlation function 
$\langle\Delta({\bf r}_i-{\bf r}_j)\rangle$. We did this by
defining the average $\langle \dots \rangle$ over a set
of compact conformations. We defined a compact conformation as
having 7  or more contacts. Reducing this number to 5 contacts
makes little difference to our results.  Only 22 out of 584 structures 
were missed and 295 folded correctly to a unique ground state. 
This is an overall success rate of over 50\%.

\section{Conclusions}
In conclusion, we have demonstrated for two simple models that we
have found a method far superior to that previously used in designing
sequences to fold to a desired structure. It is a
cumulant expansion approximation to
$\Delta F$, the difference between clamped and unclamped free energies. 
It is superior to energy minimization in several ways. 
First, it designs sequences that correctly fold into 
the desired structure, more often than energy minimization.
Second, the sequences tend to give a lower ground state degeneracy.
Third, minimizing $\Delta F$, by construction should design maximally
stable conformations. 

However the cumulant approximation used in this work is still
not perfect. As stated above for the HP model, our method
designs unique ground states in only 50\%  of the conformations
that actually have unique ground states.
Further work on improving this method is still essential
in perfecting protein design.

One exciting new approach to the reverse folding problem
has recently been proposed for the HP model~\cite{UCSF}.
A novel minimization function has been proposed that is
quite different than our $\Delta F$. It also
involves adding an additional term to the energy,
but its form its quite unlike $F_o$. Further
work connecting these two approaches is in progress.

\section{Acknowledgments}
One of us (T.K.) wishes to thank Hemant Bokil
for useful discussions. J.M.D. would like to thank
Ken Dill and Hue Sun Chan for very valuable discussions.
This work is supported by NSF grant number DMR-9419362
and acknowledgment
is made to the Donors of the Petroleum Research Fund, administered
by the American Chemical Society for partial support of this research.

\end{document}